\documentstyle[aps,prl]{revtex}
\input epsf.tex
\epsfclipon
\begin{document}
\twocolumn[\hsize\textwidth\columnwidth\hsize\csname@twocolumnfalse\endcsname

\draft
\title{Magic numbers in exotic nuclei and spin-isospin properties of $NN$
       interaction}
\author{Takaharu Otsuka$^{1,2}$,
        Rintaro Fujimoto$^{1}$, 
        Yutaka Utsuno$^{3}$, 
        B.Alex Brown$^{4}$, 
        Michio Honma$^5$, 
        and Takahiro Mizusaki$^6$}
\address{$^1$Department of Physics, University of Tokyo, Hongo, Bunkyo-ku,  
         Tokyo, 113-0033, Japan}
\address{$^2$RIKEN, Hirosawa, Wako-shi, Saitama 351-0198, Japan}
\address{$^3$Japan Atomic Energy Research Institute, Tokai, Ibaraki 
          319-1195, Japan}
\address{$^4$National Superconducting Cyclotron Laboratory, 
          Michigan State University, East Lansing, MI 48824}  
\address{$^5$Center for Mathematical Sciences, University of Aizu, Tsuruga, 
             Ikki-machi, Aizu-Wakamatsu, Fukushima 965-8580, Japan}
\address{$^6$Department of Law, Senshu University, Higashimita, 
         Tama, Kawasaki, Kanagawa, 214-8580, Japan}
\date{July 14, 2001 (PRL in press)}

\maketitle

\begin{abstract}
The magic numbers in exotic nuclei are discussed, and their novel origin is
shown to be the spin-isospin dependent part of the nucleon-nucleon interaction
in nuclei.  The importance and robustness of this mechanism is shown in 
terms of meson exchange, G-matrix and QCD theories.  In neutron-rich exotic
nuclei, magic numbers such as $N$=8, 20, {\it etc.} can disappear, 
while $N$=6, 16, {\it etc.} arise, affecting the structure of lightest 
exotic nuclei to nucleosynthesis of heavy elements.    
\end{abstract}

]


The magic number is the most fundamental quantity governing the nuclear
structure.  The nuclear shell model has been started by Mayer and Jensen 
by identifying the magic numbers and their origin \cite{MJ}.  The study of 
nuclear structure has been advanced on the basis of the shell structure
associated with the magic numbers.  This study, on the other hand, has been 
made predominantly for stable nuclei, which are on or near the 
$\beta$-stability line in the nuclear chart.  This is basically because only 
those nuclei have been accessible experimentally.  In such stable nuclei, the 
magic numbers suggested by Mayer and Jensen remain valid, and the shell 
structure can be understood well in terms of the harmonic oscillator potential
with a spin-orbit splitting.
  
Recently, studies on exotic nuclei far from the $\beta$-stability line have 
started owing to the development of radioactive nuclear beams \cite{tani}.  
The magic numbers
in such exotic nuclei can be a quite intriguing issue.  
In this Letter, we show that new magic numbers appear and some others 
disappear in moving from stable to exotic nuclei 
in a rather novel manner due to a particular part of the nucleon-nucleon 
interaction.

In order to understand underlying single-particle properties of a nucleus, 
we can make use of {\it effective (spherical) single-particle energies 
(ESPE's)},
which represent mean effects from the other nucleons on a nucleon in a 
specified single-particle orbit.  
The two-body matrix element of the interaction    
depends on the angular momentum $J$, coupled by two interacting nucleons
in orbits $j_1$ and $j_2$.  
Since we are investigating a mean effect, this $J$-dependence is averaged
out with a weight factor $(2J+1)$, and only diagonal matrix elements are
taken.  
Keeping the isospin dependence, $T$=0 or 1, the so-called monopole Hamiltonian 
is thus obtained with a matrix element \cite{monopole,mcsm-n20}:


\begin{equation} \label{mono}
V_{j_1 j_2}^{T} = \displaystyle\frac{\sum_{J}(2J+1)
< j_1 j_2 | V | j_1 j_2 >_{JT}}
{\sum_{J}(2J+1)},
\end{equation}
where $< j_1 j_2 | V | j_1' j_2' >_{JT}$ stands for the matrix element of
a two-body interaction, $V$. 

The ESPE is evaluated from this
monopole Hamiltonian as a measure of mean effects from the other nucleons.  
The normal filling configuration is used.  Note that, because the 
$J$ dependence is taken away, only the number of nucleons in each orbit 
matters.  As a natural assumption, the possible lowest isospin coupling 
is assumed for protons and neutrons in the same orbit.  
The ESPE of an {\it occupied} orbit is 
defined to be the separation energy of this orbit with the
opposite sign.  Note that the separation energy implies  
the minimum energy needed to take a nucleon out of this orbit.
The ESPE of an {\it unoccupied} orbit is 
defined to be the binding energy gain by putting a proton or neutron into  
this orbit with the opposite sign.  

\begin{figure}[tb]
 \begin{center}
 \begin{picture}(235,170)
  \put(0,0){\epsfxsize 235pt \epsfbox{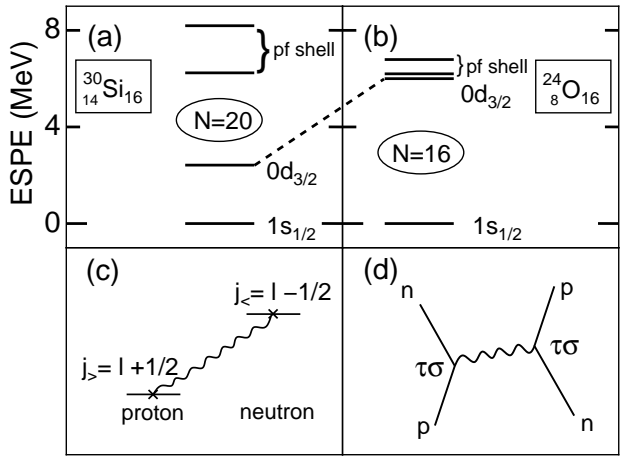}} 
 \end{picture} 
 \end{center}
 \caption{ Neutron ESPE's for (a) $^{30}$Si and (b) $^{24}$O,
           relative to $1s_{1/2}$.  The dotted line connecting (a) and (b)  
           is drawn to indicate the change of the $0d_{3/2}$ level.     
           (c) The major interaction producing the basic change between 
           (a) and (b).  (d) The process relevant to the interaction in (c).
           \label{n16_spe}}
\end{figure}

Figure~\ref{n16_spe} shows neutron ESPE's for $^{30}$Si and $^{24}$O, both of
which have $N$=16.  
The Hamiltonian and the single-particle model space are 
the same as those used in \cite{mcsm-n20}, where the structure of exotic
nuclei with $N$$\sim$20 has been successfully described within a single
framework.  The valence orbits are then $0d_{5/2,3/2}$, $1s_{1/2}$, 
$0f_{7/2}$ and $1p_{3/2}$.
 
The nucleus $^{30}$Si has six valence protons in the
$sd$ shell and is a stable nucleus, 
while $^{24}$O has no valence proton and is a neutron-rich exotic nucleus.
For $^{30}$Si, the neutron $0d_{3/2}$ and $1s_{1/2}$ are
rather close to each other  (see Fig.~\ref{n16_spe} (a)).  
For $^{24}$O, as shown in Fig.~\ref{n16_spe} (b), 
the $0d_{3/2}$ is lying much higher and is quite close to the $pf$ shell, 
giving rise to a large gap ($\sim$6 MeV) between $0d_{3/2}$ and $1s_{1/2}$ 
\cite{mexico}.  On the other hand, for the stable nucleus $^{30}$Si,
a considerable gap ($\sim$4 MeV) is created between the $0d_{3/2}$ 
and the $pf$ shell (See Fig.~\ref{n16_spe} (a)). 
Thus, the $N$=20 magic structure is evident for 
$^{30}$Si, whereas the $N$=16 magic number arises in $^{24}$O.  
In $^{24}$O, the $0d_{3/2}$
is lying higher reflecting the large spin-orbit splitting which is basically
the same as that for $^{17}$O.  Although this high-lying $0d_{3/2}$ orbit 
is not so relevant to the ground state of lighter O isotopes, it should 
affect binding energies of nuclei around $^{24}$O.  
Such an anomaly was pointed out by Ozawa {\it et al.} \cite{ozawa}
in observed binding energy systematics.

The dramatic change of ESPE's from $^{24}$O to
$^{30}$Si is primarily due to the strongly attractive interaction 
between a proton in $0d_{5/2}$ and a neutron in $0d_{3/2}$.
A schematic picture on this point is shown in Fig.~\ref{n16_spe} (c)
for the general cases comprised of a pair of orbits 
$j_> = l+1/2$ and $j_< = l-1/2$ with $l$ being
the orbital angular momentum.
Note that $j_>$ and $j_<$ are nothing but spin-orbit coupling partners.  
The present case corresponds to $l$=2.  
As one moves from $^{24}$O to $^{30}$Si, 
six valence protons are put into the $0d_{5/2}$ orbit.
Consequently, due to the strong attraction shown in Fig.~\ref{n16_spe} (c), 
a neutron in $0d_{3/2}$ is more bound in $^{30}$Si, and 
its neutron $0d_{3/2}$ ESPE becomes so low 
as compared to that in $^{24}$O where such attraction is absent.

The process illustrated in Fig.~\ref{n16_spe} (d) produces  
the attractive interaction in Fig.~\ref{n16_spe} (c). 
The nucleon-nucleon ($NN$) interaction in this process is written as  
\begin{equation} \label{ts}
V_{\tau\sigma} = \tau \cdot \tau \,\, \sigma \cdot \sigma \, f_{\tau\sigma}(r).
\end{equation}
Here, the symbol ``$\cdot$'' denotes a scalar product, 
$\tau$ and $\sigma$ stand for isospin and spin operators, respectively,
$r$ implies the distance between two interacting nucleons, and 
$f_{\tau\sigma}$ is a function of $r$.  In the long range (or no r-dependence) 
limit of $f_{\tau\sigma}(r)$,
the interaction in eq.(\ref{ts}) can couple only a pair of orbits with the same
orbital angular momentum $l$, which are nothing but $j_>$ or $j_<$.  

The $\sigma$ operator couples $j_>$ to $j_<$ (and vice versa) more
strongly than $j_>$ to $j_>$ or $j_<$ to $j_<$.  Therefore, the spin flip
process is more favored in the vertexes in Fig.~\ref{n16_spe} (d), 
while spin non-flip contributions certainly exist.  
The same mathematical mechanism 
works for isospin: the $\tau$ operator favors charge exchange processes.  
Combining these two properties,
$V_{\tau\sigma}$ produces large matrix elements for 
the spin-flip isospin-flip processes: 
proton in $j_>$ $\rightarrow$ neutron in $j_<$ and vice versa.  
This gives rise to the interaction in Fig.~\ref{n16_spe} (c)
with a strongly attractive monopole term for the appropriate sign of
$V_{\tau\sigma}$.  
This feature is a general one and is maintained with $f_{\tau\sigma}(r)$ in 
eq.(\ref{ts}) with reasonable $r$ dependences.


In stable nuclei with $N$$\sim$$Z$ with
ample occupancy of the $j_>$ orbit in the valence shell, the proton 
(neutron) $j_<$ 
orbit is lowered by neutrons (protons) in the $j_>$ orbit.  
In exotic nuclei, this lowering can be absent, and then the $j_<$ orbit is 
located rather high, not far from the upper shell.  
In this sense, the proton-neutron 
$j_>$-$j_<$ interaction enlarges a gap between major 
shells for stable nuclei with proper occupancy of relevant orbits.  
This interaction has been known, for instance \cite{pittel,casten}, to play 
important roles also in other issues, {\it e.g.}, the onset of the deformation.

The origin of the strong   
$V_{\tau\sigma}$ is quite clear.  The One-Boson-Exchange-Potentials
(OBEP) for $\pi$ and $\rho$ mesons have this type of terms as major
contributions.  While the OBEP is one of major parts of the effective
$NN$ interaction, the effective $NN$ interaction in nuclei can be
provided by the G-matrix calculation with core polarization corrections
and other various effects.
Such effective $NN$ interaction will be called simply G-matrix interaction
for brevity.
The G-matrix interaction should maintain the basic features of meson exchange 
processes, and, in fact, existing G-matrix interactions generally have 
quite large matrix elements for the cases shown in Fig.~\ref{n16_spe} (c), 
including strongly attractive monopole terms \cite{morten}.  

We would like to point out that 
the $1/N_c$ expansion of QCD by Kaplan and Manohar 
indicates that $V_{\tau\sigma}$ is one of three leading terms 
of the long-range part of the $NN$ interaction \cite{kaplan}.  
Since the next order of this expansion is smaller by a factor $(1/N_c)^2$, 
the leading terms should have rather distinct significance.
One of the other two leading terms in the $1/N_c$ expansion \cite{kaplan} 
is a central force,
\begin{equation} \label{central}
V_0 = f_0 \, (r),
\end{equation}
where $f_0$ is a function of $r$.  The last leading term is a tensor force.  
We shall come back to these forces later.

\begin{figure}[tb]
 \begin{center}
 \begin{picture}(150,130)
  \put(0,0){\epsfxsize 140pt \epsfbox{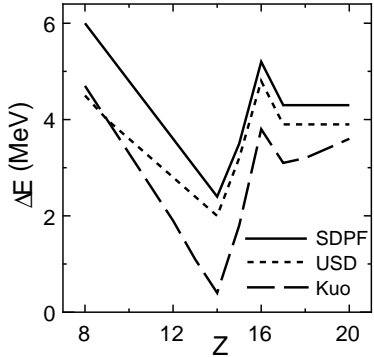}} 
 \end{picture} 
 \end{center}
 \caption{ Effective $1s_{1/2}$-$0d_{3/2}$ gap in
           $N$=16 isotones as a function of $Z$.      
           Shell model Hamiltonians, SDPF, USD and ``Kuo'' are used. 
           See the text.
          \label{sd3_gap}}
\end{figure}

Figure~\ref{sd3_gap} shows the effective $0d_{3/2}$\,-\,$1s_{1/2}$ gap, 
{\it i.e.}, the difference between ESPE's of these orbits, 
in $N$=16 isotones with $Z$=8$\sim$20 
for three interactions: ``Kuo'' means 
a G-matrix interaction for the $sd$ shell calculated by Kuo \cite{kuo},   
and USD was obtained by adding empirical modifications to ``Kuo''
\cite{usd}.  
The present shell-model interaction is denoted SDPF hereafter, and 
its $sd$-shell part is nothing but USD with small changes 
\cite{mcsm-n20}.  Steep decrease of this gap is found 
in all cases, as $Z$ departs 
from 8 to 14.  In other words, a magic structure can be formed around $Z$=8, 
but it should 
disappear quickly as $Z$ deviates from 8 because the gap decreases very fast.  
The slope of this sharp drop is determined by  
$V_{0d_{5/2} 0d_{3/2}}^{T=0,1}$ in eq.~(\ref{mono}), where  
the dominant contribution is from $T$=0.
Note that $V_{0d_{5/2} 0d_{3/2}}^{T=0}$ is most attractive among 
all $V_{j_1 j_2}^{T}$'s in the $sd$ shell for each of the three interactions.  
In SDPF, for instance, its magnitude 
exceeds by $\sim$1 MeV that of the second most attractive one.  

We shall now discuss briefly the present gap in other approaches.
The gap can be calculated from the Woods-Saxon 
potential which is a standard modeling of
single-particle structure for stable nuclei.  The resultant gap is 
rather flat, and is about half of the SDPF value for $Z$=8.
With Skyrme Hartree-Fock (HF) interactions, the gap changes more  
smoothly and gradually, too.    

The occupation number of the neutron $1s_{1/2}$ is calculated 
by Monte Carlo Shell Model \cite{qmcd-ph3} 
with full configuration mixing for the nuclei shown in Fig.~\ref{sd3_gap}.
It is nearly two for $^{24}$O 
as expected for a magic nucleus, but decreases sharply
as $Z$ increases.  It remains smaller ($<$ 1.5) in the middle region around 
$Z$=14, and finally goes up again for $Z$$\sim$20.  
This means that the $N$=16 magic structure is broken in the middle region 
of the proton $sd$ shell, where
deformation effects also contribute to the breaking.
The $N$=16 magic number is thus quite valid at both ends.  
It is of interest that the gap becomes large again for larger $Z$, 
due to other monopole components.

We now turn to exotic nuclei with $N$$\sim$20.
The ESPE has been evaluated for them in \cite{mcsm-n20}. 
The small effective gap between $0d_{3/2}$ and the $pf$ shell for 
neutrons is obtained, and is found to play essential roles for various
anomalous features.
This small gap is nothing but what we have seen for $^{24}$O in 
Fig.~\ref{n16_spe} (b).
Thus, the disappearance of $N$=20 magic structure in $Z$=9$\sim$14
exotic nuclei and the appearance of the new magic structure in $^{24}$O
have the same origin.

\begin{figure}[tb]
 \begin{center}
 \begin{picture}(210,235)
  \put(0,0){\epsfxsize 210pt \epsfbox{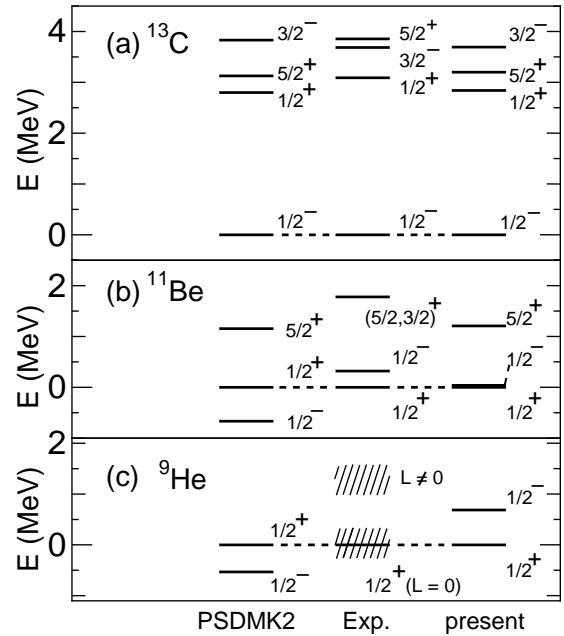}} 
 \end{picture} 
 \end{center}
 \caption{ Energy levels for (a) $^{13}$C, (a) $^{11}$Be and (c) $^{9}$He,
           relative to the state corresponding to the experimental ground 
           state.   The results by the PSDMK2 and present Hamiltonians
           are compared 
           to experiments.  Hatched area indicate continuum states with
           resonance(-like) structures.
           \label{n7_level}}
\end{figure}

A very similar mechanism works for $p$-shell nuclei.
We consider the structure of a stable nucleus $^{13}$C, and exotic 
nuclei $^{11}$Be and $^{9}$He, all of which have $N$=7.
Shell model calculations are performed with the PSDMK2 
Hamiltonian \cite{GTLi} in the $p$+$sd$ shell model space 
on top of the $^4$He core.
Figure~\ref{n7_level} indicates that 
the experimental levels are reproduced well 
for $^{13}$C \cite{table}, whereas notable deviations are
found in $^{11}$Be and $^{9}$He \cite{table,9He}.  
Here, for experimental levels in the continuum,  
their resonance(-like) energies are compared to the shell-model
results.

The $p$-shell part of PSDMK2 is the Cohen-Kurath (CK) 
Hamiltonian \cite{CK}.
The single-particle energies of $0p_{3/2}$ and $0p_{1/2}$ are 1.38 
and 1.68 MeV, respectively, in PSDMK2.
These energies correspond to the observed spectra of 
$^5$He: 1.15 for $0p_{3/2}$ and $\sim$5 MeV for $0p_{1/2}$.
We use these observed values as single particle energies, 
while $0p_{1/2}$ is quite low in PSDMK2.
As compared to the G-matrix interaction \cite{morten}, 
the interaction in Fig.~\ref{n16_spe} (c) for the $p$-shell 
is too weak in CK, and is enlarged by 
shifting all $<0p_{3/2}\, 0p_{1/2}|V|0p_{3/2}\, 0p_{1/2}>_{J,T=0}$'s 
by -2 MeV independently of $J$. 
Thus, the Hamiltonian is modified only for three parameters.
Since we are focusing on basic trends and would like to
eliminate spurious center-of-mass components, all shell model calculations
for $N$=7 isotones 
are made in the so-called 0$\hbar\omega$ or 1$\hbar\omega$ space.  
Figure ~\ref{n7_level} (a) indicates that the levels of $^{13}$C calculated 
from the
present Hamiltonian are similar to the ones obtained from PSDMK2,
because the higher-lying $0p_{1/2}$ is pulled down by $V$ as discussed above.

The ground state of $^{11}$Be and $^{9}$He
is known for the {\it inversion} between 1/2$^+$ and 1/2$^-$ \cite{tu}. 
The present calculation reproduces this inversion for both nuclei, 
whereas the PSDMK2 fails in either case.  
With the present Hamiltonian, neutron effective $1s_{1/2}$\,-\,$0p_{1/2}$ gap  
decreases from 8.6 MeV for $^{13}$C down to 0.8 MeV for $^{9}$He.
Dynamical correlations finally invert
1/2$^+$ and 1/2$^-$ eigenstates \cite{dyn}.
Thus, we achieve a reasonable description of stable and exotic nuclei with 
$N$=7.  Although the result appears promising, we have investigated only 
possible improvements, and further studies
are needed for an overall 
description of $p$-$sd$ nuclei.  For instance, in comparison to experiments, 
the present interaction gives a 3/2$^-$ state in $^{15}$O $\sim$2 MeV too 
high \cite{table}, and $^9$He is too unbound relative to $^8$He by 
$\sim$4.5 MeV \cite{9He}.  The latter may be related to coupling to 
continuum and/or to nuclear-size dependence of two-body interaction.

The neutron $0p_{1/2}$ orbit becomes higher as the nucleus loses
protons in its spin-flip partner $0p_{3/2}$.  
In nuclei such as He, Li and Be, 
the $N$=8 magic structure then disappears.  In some cases,  
$N$=6 becomes magic: for instance, bound $^8$He and unbound $^{9}$He
are obtained, similarly to bound $^{24}$O and unbound $^{25}$O.  

As the neutron $0p_{1/2}$ is shifted quite high in He, Li and Be, 
this orbit becomes close to the $1s_{1/2}$ which may come down 
due to halo structure.  This is a situation like the Efimov state
\cite{Efimov}.  
The pairing between these two orbits, including contribution from the
tensor force, may stabilize the two-neutron halo and provide us with 
bound $^{11}$Li, {\it etc.}
Thus, the present mechanism plays a crucial role
for the structure of dripline nuclei.  Without this mechanism, it is
usually difficult to make $0p_{1/2}$ and $1s_{1/2}$ so close to each
other, except for an empirical shift of the single-particle energy.

Moving back to heavier nuclei, from the strong interaction in 
Fig.~\ref{n16_spe} (c), we can predict other magic numbers, for instance,
$N$=34 associated with the $0f_{7/2}$\,-\,$0f_{5/2}$ interaction.  In heavier
nuclei, $0g_{7/2}$, $0h_{9/2}$, {\it etc.} are shifted upward in 
neutron-rich exotic nuclei, disturbing the magic numbers $N$=82, 126, 
{\it etc.}   It is of interest how the $r$-process of nucleosynthesis
is affected by it.   

In conclusion, we showed how magic numbers are changed in 
nuclei far from the $\beta$-stability line: $N$=6, 16, 34, {\it etc.} 
can become magic numbers in neutron-rich exotic 
nuclei, while usual magic numbers, $N$=8, 20, 40, {\it etc.}, may disappear.  
Since such changes occur as results of the nuclear force, there is isospin
symmetry that similar changes occur for the same $Z$ values in mirror nuclei.
The mechanism of this change can be explained by the strong  
$V_{\tau\sigma}$ interaction which has robust origins in
OBEP, G-matrix and QCD.  In fact, simple structure such as magic numbers
should have a simple and sound basis.  
Since it is unlikely that a mean central potential can 
simulate most effects of $V_{\tau\sigma}$, we should treat $V_{\tau\sigma}$ 
rather explicitly.   It is nice to build a bridge between very basic feature 
of exotic nuclei and the basic theory of hadrons, QCD.  
In existing Skyrme HF calculations except for those with Gogny 
force, effects of $V_{\tau\sigma}$ 
may not be well enough included,
because the interaction is truncated to be of $\delta$-function type. 
The Relativistic
Mean Field calculations must include pion degrees of freedom to be
consistent with $V_{\tau\sigma}$.  Thus, the importance of $V_{\tau\sigma}$
opens new directions for mean field theories of nuclei.
In addition, $f_0(r)$ in eq.(\ref{central}) and $f_{\tau\sigma}(r)$ 
in eq.(\ref{ts}) should have different ranges.  Since the latter has
smaller contributions in exotic nuclei, this difference should produce very
interesting effects on nuclear size.
Loose-binding or continuum effects are important in some exotic 
nuclei.  By combining such effects with those discussed in this Letter
one may draw a more complete picture for the structure of exotic nuclei. 
Finally, we would like to mention once more that the $V_{\tau\sigma}$ 
interaction should produce 
large, simple and robust effects on various properties, and may 
change the landscape of nuclei far from the $\beta$-stability line 
in the nuclear chart.    

One of the authors (T.O.) acknowledges the Institute for Nuclear Theory
for inviting him to the INT-00-3 program.
He is grateful to Prof. K. Yazaki for valuable comment. 
We thank Prof. A. Gelberg for reading the manuscript.
The MCSM calculations were performed by the Alphleet computer 
in RIKEN. The conventional shell-model calculations were 
made by the code {\sc oxbash} \cite{oxbash}. 
This work was supported in part by Grant-in-Aid for Scientific Research
(A)(2) (10304019) from the Ministry of Education, Science and Culture.

\end{document}